\documentclass[%
 reprint,
superscriptaddress,
 amsmath,amssymb,
 aps,
pra,
floatfix,
]{revtex4-2}

\usepackage{graphicx}
\usepackage{dcolumn}
\usepackage{bm}

\usepackage{lineno}
\usepackage{graphicx}
\usepackage{color,soul}
\usepackage{outlines} 
\usepackage{setspace}

\usepackage{listings}  
\begin{document}

\preprint{APS/123-QED}

\title{Laser cooling $^{88}$Sr to microkelvin temperature with an integrated-photonics system}

\author{Andrew R. Ferdinand}
 \affiliation{Time and Frequency Division, National Institute of Standards and Technology, Boulder, Colorado, 80305, USA}
 \affiliation{Department of Physics, University of Colorado, Boulder, Colorado, 80309, USA}
 
\author{Zheng Luo}
\affiliation{Time and Frequency Division, National Institute of Standards and Technology, Boulder, Colorado, 80305, USA}
\affiliation{Department of Physics, University of Colorado, Boulder, Colorado, 80309, USA}

\author{Sindhu Jammi}
\affiliation{Time and Frequency Division, National Institute of Standards and Technology, Boulder, Colorado, 80305, USA}
\affiliation{Department of Physics, University of Colorado, Boulder, Colorado, 80309, USA}

\author{Zachary Newman}
\affiliation{Octave Photonics, Louisville, Colorado, 80027, USA}

\author{Grisha Spektor}
\affiliation{Time and Frequency Division, National Institute of Standards and Technology, Boulder, Colorado, 80305, USA}
\affiliation{Department of Physics, University of Colorado, Boulder, Colorado, 80309, USA}
\affiliation{Octave Photonics, Louisville, Colorado, 80027, USA}

\author{Okan Koksal}
\affiliation{Microsystems and Nanotechnology Division, National Institute of Standards and Technology, Gaithersburg, Maryland, 20899, USA}

\author{Parth B. Patel}
\affiliation{Vector Atomic, Inc., Pleasanton, California 94588, USA}

\author{Daniel Sheredy}
\affiliation{Vector Atomic, Inc., Pleasanton, California 94588, USA}

\author{William Lunden}
\affiliation{Vector Atomic, Inc., Pleasanton, California 94588, USA}

\author{Akash Rakholia}
\affiliation{Vector Atomic, Inc., Pleasanton, California 94588, USA}

\author{Travis C. Briles}
\affiliation{Time and Frequency Division, National Institute of Standards and Technology, Boulder, Colorado, 80305, USA}

\author{Wenqi Zhu}
\affiliation{Microsystems and Nanotechnology Division, National Institute of Standards and Technology, Gaithersburg, Maryland, 20899, USA}

\author{Martin M. Boyd}
\affiliation{Vector Atomic, Inc., Pleasanton, California 94588, USA}

\author{Amit Agrawal}
\affiliation{Microsystems and Nanotechnology Division, National Institute of Standards and Technology, Gaithersburg, Maryland, 20899, USA}

\author{Scott B. Papp}
\email{scott.papp@nist.gov}
\affiliation{Time and Frequency Division, National Institute of Standards and Technology, Boulder, Colorado, 80305, USA}
\affiliation{Department of Physics, University of Colorado, Boulder, Colorado, 80309, USA}

\date{\today}

\begin{abstract}
We report on experiments generating a magneto-optical trap (MOT) of 88-strontium ($^{88}$Sr) atoms at microkelvin temperature, using integrated-photonics devices. With metasurface optics integrated on a fused-silica substrate, we generate six-beam, circularly polarized, counter-propagating MOTs on the blue broad-line, 461 nm, and red narrow-line, 689 nm, Sr cooling transitions without bulk optics. By use of a diverging beam configuration, we create up to 10 mm diameter MOT beams at the trapping location. To frequency stabilize and linewidth narrow the cooling lasers, we use fiber-packaged, integrated nonlinear waveguides to spectrally broaden a frequency comb. The ultra-coherent supercontinuum of the waveguides covers 650 nm to 2500 nm, enabling phase locks of the cooling lasers to hertz level linewidth. Our work highlights the possibility to simplify the preparation of an ultracold $^{88}$Sr gas for an optical-lattice clock with photonic devices. By implementing a timing sequence for  control of the MOT lasers and the quadrupole magnetic-field gradient, we collect atoms directly from a thermal beam into the blue MOT and continuously cool into a red MOT with dynamic detuning and intensity control. There, the red MOT temperature is as low as $2~\mu$K and the overall transfer efficiency up to 16 $\%$. We characterize this sequence, including an intermediate red MOT with modulated detuning.
Our experiments demonstrate an integrated photonics system capable of cooling alkaline-earth gases to  microkelvin temperature with sufficient transfer efficiencies for adoption in scalable optical clocks and quantum sensors.
\end{abstract}

\maketitle

Access to ultracold samples of alkaline-earth atomic species promotes development in optical-lattice clocks~\cite{Ludlow2015OpticalClocks}, searches for fundamental physics~\cite{Safronova18RMPsearch}, exploration of quantum matter~\cite{Rey13scienceQuantumManyBody}, and quantum simulation and computation~\cite{Kaufman21NatPhotRevQuantumScienceColdAtoms}. Species like Sr offer a variety and richness of atomic transitions that facilitate numerous physical interactions and controls with optical fields. A broad linewidth transition at 461 nm enables robust laser cooling from a vapor or atomic beam source, referred to as a blue magneto-optical trap (MOT). The spectrally narrow intercombination transitions allow for further cooling to microkelvin temperature at 689 nm, referred to as a red MOT, and a provide high-precision, high-stability frequency reference at 698 nm for optical clocks~\cite{Nicholson15_srII_systematic_eval} and quantum information~\cite{Kaufman21NatPhotRevQuantumScienceColdAtoms}.

The complexity of controlling and manipulating Sr gases leads to challenges in the development of systems suitable for applications beyond laboratory experiments. Employing Sr MOTs requires addressing the atoms with a complex, three-dimensional beam geometry with both 461 nm and 689 nm laser wavelengths in a dynamic experimental sequence. Moreover, experiments require a source of Sr vapor and an ultrahigh vacuum environment. Laboratory type strontium systems generally implement a MOT beam geometry with manually aligned, table-top, bulk optic systems. Cooling on the narrow-line transition requires precise frequency stabilization, which is complicated by low vapor pressure in room-temperature cells. Furthermore, the light must be of narrow enough linewidth to address the atomic transitions. Laboratory systems typically consist of high finesse bulk optical-reference cavities and complex arrangements of highly nonlinear fiber and nonlinear crystals. The incompatibility of specialized and bulky laboratory solutions for transportable cold-atom systems motivates development of alternative technologies, especially with integrated photonics.

Here, we report a system to create a sample of $^{88}$Sr atoms laser-cooled to microkelvin temperature with integrated- photonics devices. Our system uses metasurface (MS) optics integrated on a common substrate to generate a complete, multi-wavelength, three-dimensional set of MOT beams without the use of  bulk optics. The metasurface-optics system implements beam routing, polarization, pointing, and divergence control for both sets of MOT beams. We frequency stabilize the cooling lasers to a frequency-comb supercontinuum generated with fiber-packaged integrated nonlinear waveguides. The supercontinuum provides high-power modes at 689 nm, 698 nm, 813 nm, and 922 nm to reference all the lasers needed in a Sr optical clock. The reduced complexity of our integrated-photonics approach eliminates barriers to production of robust ultracold atom samples for applications outside of specialized laboratories.
\begin{figure}[t!]
\centering
\includegraphics[width=\linewidth, trim=6 2 4 2,clip]{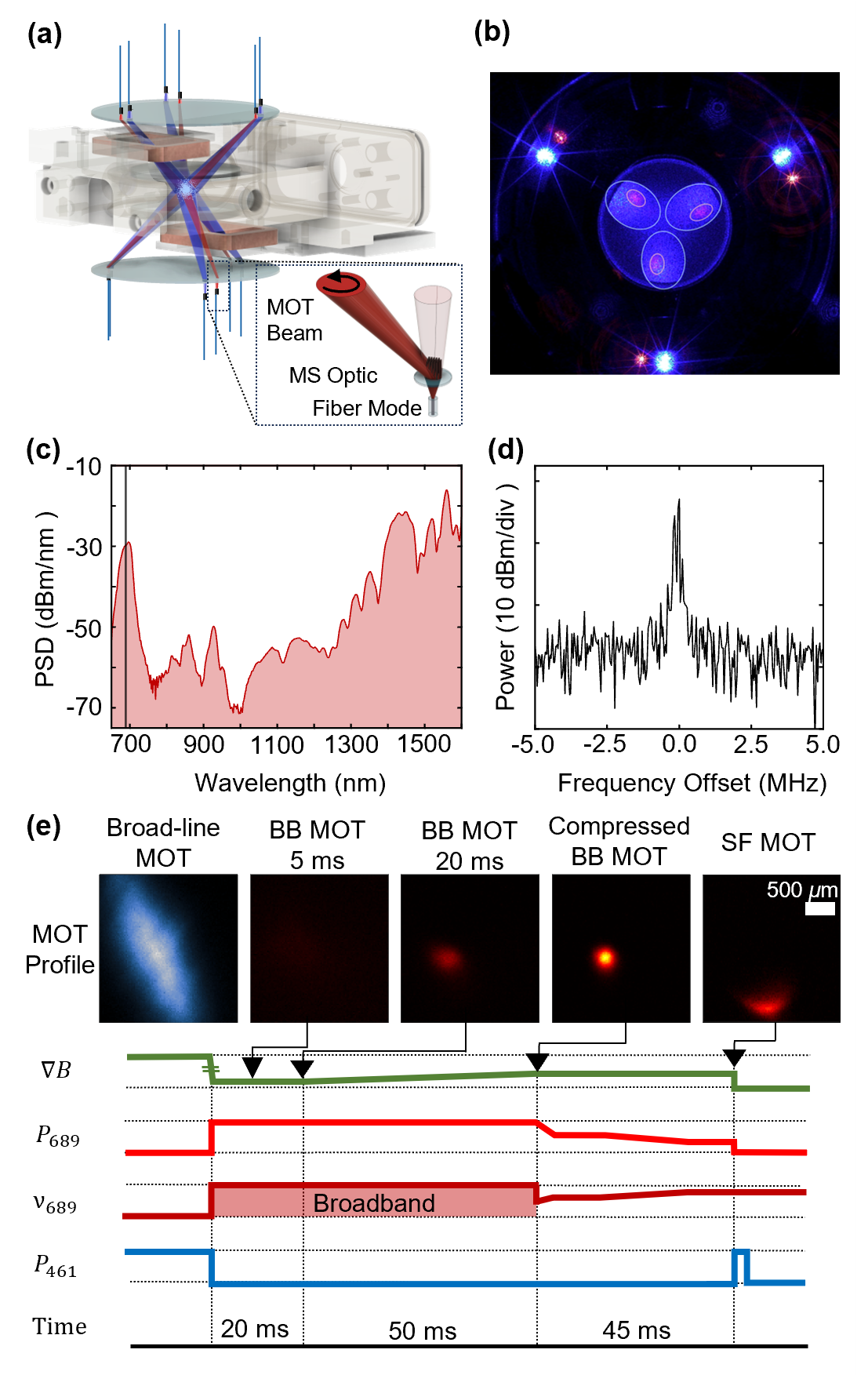}
\caption[\parbox{3.25in}]{
Sr system with integrated photonics.   
(a) Metasurface-based system. Inset depicts the generation of a MOT beam with a multifunctional MS optic; 
(b) Multi-wavelength MOT beam cross section $7.5~$mm from trap center; 
(c) Supercontinuum power spectral density (PSD) with dispersive wave at target 689 nm wavelength.
(d) Heterodyne beat of 689 nm cooling light and supercontinuum
(e) Experimental sequence with select false color MOT images, including broadband (BB) and single frequency (SF) narrow-line MOTs.
\label{fig:fig1}}
\end{figure}

Figure 1 introduces the apparatus we use, including MS optics, magnetic-field coils, waveguide supercontinuum, vacuum chamber, and the timing sequence to laser cool $^{88}$Sr. We create the twelve-beam, free-space configuration of 461 nm and 689 nm MOT beams with metasurface optics, which are integrated on two fused-silica substrates \cite{Jammi2024Three-dimensionalCooling}. By illuminating each MS optic with light directly expanding from a polarization-maintaining fiber, we simultaneously engineer the optical phase profile for near arbitrary pointing angle, beam divergence, and polarization; see Fig. 1(a). This integrated approach simplifies formation of MOT beams because each metasurface optic performs the function of mirrors, lenses, beam splitters, and wave plates. To implement the MS system with Sr atoms, we have developed a flattened, titanium ultrahigh vacuum chamber with large viewports to accept the beams from the MS substrates. Moreover, we load Sr atoms into the MOT from a collimated, thermal atom beam, Doppler slowed on the $^{1}$S$_0 \rightarrow ^{1}$P$_1$ transition~\cite{Ferdinand24Infrastructure}

Our propagation phase MSs~\cite{Agrawal20_prl_MS} are arrays of sub-wavelength TiO$_2$ nanopillars that impart a phase shift on the optical beam. The phase shift varies, depending on the dimensions of the intercepted nanopillar. We numerically simulate these phase shifts for various pillar geometries and collect them into a library.  Then, we position pillars at locations where the imparted phase shift matches that required by the desired phase profile. The twelve metasurfaces are partitioned onto two three-inch fused-silica wafers that are placed symmetrically about our compact vacuum chamber, using mechanical fixtures.  The MS optics deflect each beam at 45 degrees toward the trap center.   We adjust the sizes of the 461 nm and 689 nm trapping beams at the trap center by varying the effective focal length through the quadratic term in the phase profile or by changing the mode diameter at the input to the MS. For the first stage of laser cooling, we choose a relatively large beam diameter at the atoms ($d=1~$cm) to create a large trap volume and maximize the total trapped atom number~\cite{Foot92RadiationForceMOT}.  For the second stage of laser cooling, we choose a smaller beam diameter of $d=3~$ mm, which is dictated largely by our available laser power and the laser intensity required to effectively cool from our millikelvin blue MOT temperature in the red MOT.  

An essential feature of our system is the integration of MSs for multiwavelength laser cooling on a single substrate.  Figure 1(b) presents a horizontal cross-section of the 461 nm and 689 nm beams generated by a single substrate at a distance $7.5~$mm below the trap center.  The outlines of the beams are digitally emphasized for clarity.  The MSs are arranged with a minimal azimuthal offset of 10$^{\circ}$ which is crucial for maintaining alignment with the magnetic field for proper operation of our MOT at both wavelengths.  This clocking angle can be seen in the residual light from the fibers that is transmitted through the wafer but undeflected by the MSs; see outer perimeter of Fig. 1(b).   

We frequency stabilize the red and blue MOT lasers to a frequency-comb supercontinuum, generated with fiber-coupled, integrated nonlinear waveguide modules. Dispersion engineering via the waveguide geometry allows us to tailor the generated supercontinuum spectra for a high power spectral density (PSD) in target spectral bands, ie. 700 nm and 922 nm. The modules are composed of a 12 mm long tantala waveguide on a silicon chip~\cite{Lamee2020NanophotonicNm} with $800$ nm thickness, $1800~$nm waveguide width, and silicon oxide top cladding that can be selectively removed from a portion of the waveguide to facilitate group-velocity dispersion engineering. We create inverse tapers at the chip edge with 180 nm critical dimension to enable low insertion loss packaging to polarization maintaining optical fiber. 
An amplified 1550 nm mode-locked fiber laser seeds the supercontinuum generation process. This mode-locked laser is carrier-envelope offset locked with a separate tantala waveguide and PPLN integrated module. We phase lock the mode-locked laser offset and repetition frequencies with respect to a hydrogen maser reference and an ultrastable 1550 nm optical reference cavity, respectively. 
Our module to generate supercontinuum for locking the narrow-line cooling laser creates a dispersive wave with $-30~$dBm/nm PSD at the target 689 nm wavelength, as seen in Fig.~\ref{fig:fig1}~(c).
To phase lock the red and blue MOT lasers, we obtain heterodyne beats with the supercontinuum and implement standard electronic phase locked loops.
An example unlocked heterodyne beat of supercontinuum and our 689 nm laser is given in Fig ~\ref{fig:fig1}~(d).
Out of loop measurements with an independent frequency comb indicate hertz level linewidth of the locked laser, sufficient for cooling on the $7.5~$kHz linewidth atomic intercombination transition of Sr. This additionally indicates the coherence of our supercontinuum is maintained spanning over a octave and into the visible wavelength range, motivating its use for laser stabilization in applications beyond specialized laboratory settings.

Achieving microkelvin temperature with Sr requires addressing both cooling transitions in a dynamic experimental sequence; see Fig.~\ref{fig:fig1}(e). 
In the first phase of the sequence, we load and cool atoms in the blue MOT from a thermal atom beam from a conventional oven, assisted by our Doppler slowing. The magnetic-field gradient along the axis between the coils $\nabla B$ is 50~G/cm. We repump atoms that decay from the excited state into the $^{3}P_J$ manifold with 707 nm and 679 nm lasers. The blue MOT cools atoms to millikelvin temperature, fundamentally limited by the Doppler limit $T_D = \hbar\Gamma/2k_{\text{B}}~= 0.7~$mK~\cite{Loftus2004NarrowCrystals}, where $\hbar$ is reduced Planck's constant, $k_{\text{B}}$ is Boltzmann's constant, and $\Gamma$ is the transition's natural linewidth. We transfer the atoms to the red MOT for further cooling on the $7.5~$kHz intercombination transition. The much lower Doppler limit of $T_D = 0.2~\mu$K is important for lattice loading and precision spectroscopy. However, the smaller linewidth and larger \textit{g}-factor necessitate dynamic control of the magnetic-field gradient, laser power, and laser frequency. During the first period of cooling on the narrow-line transition, frequency modulation of the 689 nm cooling light allows the laser to address all of the atomic velocities comprising the millikelvin temperature blue MOT, and we refer to this as the broadband (BB) MOT. 
Reducing the magnetic field gradient to $\nabla B~= 5~$G/cm optimizes scattering and improves the transfer efficiency into the BB MOT. After $20$ ms BB cooling in the lower magnetic-field gradient, we ramp the gradient to $10~$ G/cm over 50 ms, spatially compressing the atom cloud to the final BB MOT. After the BB MOT, we extinguish the frequency modulation to implement a single frequency (SF) final cooling stage. During this cooling stage, we ramp down the cooling laser intensity and detuning over 45 ms to reduce the scattering rate and further compress the atom cloud. This step reduces temperature and increases phase-space density~\cite{Blatt2019FastDenseMOT}. At sufficiently low temperature, laser intensity, and spatial confinement, the gravitational force on the atoms becomes a non-negligible contributor to the cooling potential. Therefore, the MOT shape evolves to an apparent crescent as the atoms settle near the bottom of the ellipsoidal potential~\cite{Loftus2004NarrowCrystals} ; see the MOT profile panel in Fig.~\ref{fig:fig1}~(e). In the remainder of the paper, we present characterizations of our BB MOT and SF MOTs.

\begin{figure}[ht!]
\centering
\includegraphics[width=0.9\linewidth, trim=7 0 4 4,clip]{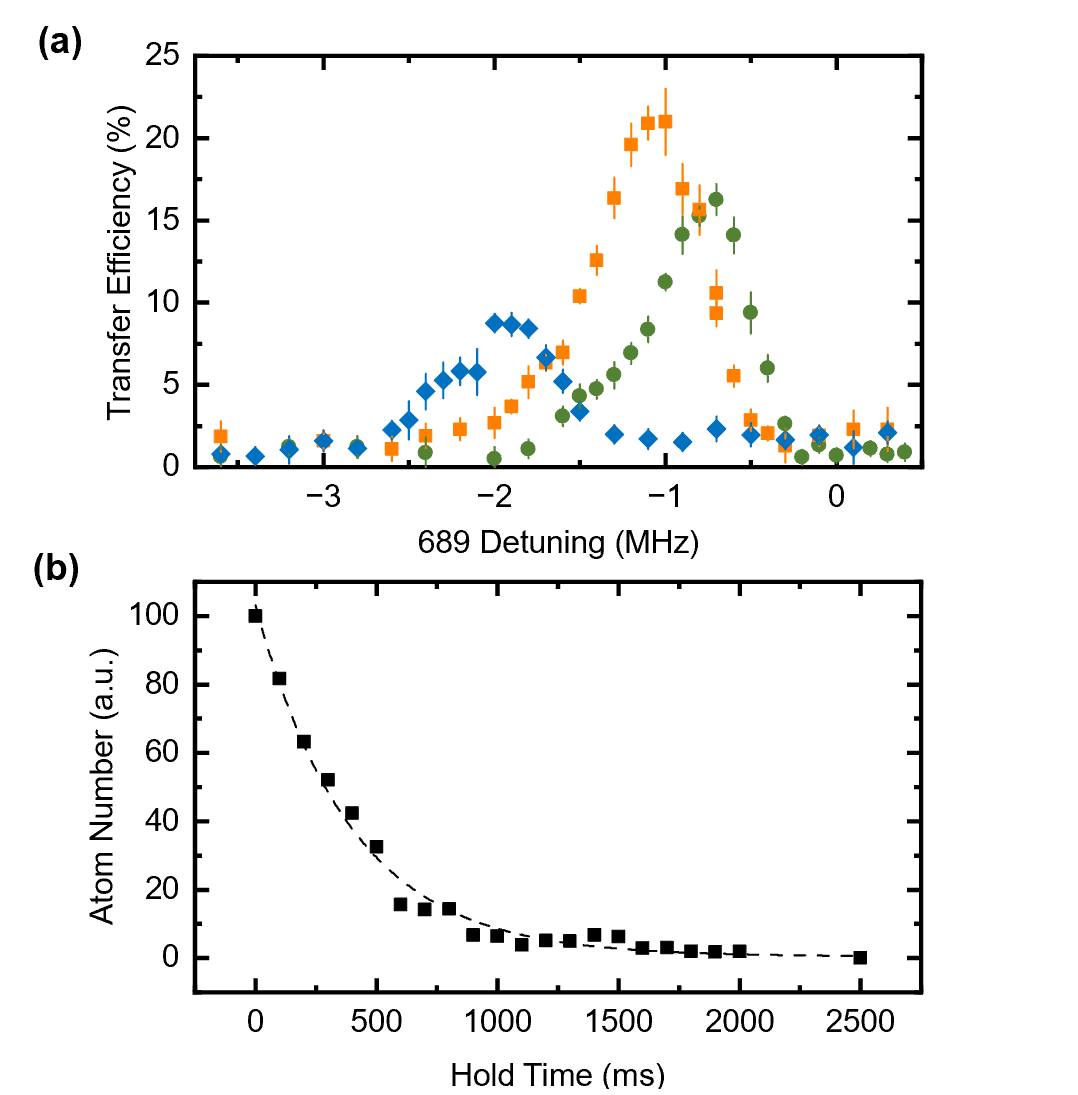}
\caption{BB MOT Characterization. (a) Transfer efficiency with 1-$\sigma$ error bars versus detuning for three values of modulation frequency range $\delta_{\text{FM}}$ : $880$ kHz (green circles), $1760$ kHz (orange squares), and $3520$ kHz (blue diamonds). Bars indicate the standard deviation. (b) Atom number versus time at the end of the BB MOT sequence. 
\label{fig:fig2}
}
\end{figure}

Figure \ref{fig:fig2} presents characterization of our BB MOT. The transfer efficiency from the blue MOT to the BB MOT depends on the spectral parameters of the 689 nm laser, including the detuning from the atomic transition and the laser linewidth broadening from frequency modulation. Effectively configured spectral parameters allow the cooling light to address the full velocity distribution of atoms in the blue MOT while maintaining a sufficient PSD to trap the atoms. In our experiments, we modulate the laser at $f_{\text{mod}} = 50$ kHz, creating sidebands across a controllable frequency range, $\delta_{\text{FM}}$, such that the laser addresses each accessible atomic velocity class once within an atomic lifetime $\tau = \Gamma^{-1}$.  In Fig.~\ref{fig:fig2}~(a), we present BB MOT transfer efficiency as a function of detuning for three values of $\delta_{\text{FM}}$: 880 kHz (green), 1760 kHz (orange), and 3520 kHz (blue).  The data exhibits similar relationships between transfer efficiency, detuning, and $\delta_{\text{FM}}$. The maximum transfer efficiency occurs when the detuning and $\delta_{\text{FM}}$ are set such that the maximum frequency sampled by the laser is a slightly lower frequency than the atomic resonance. We achieve a maximum transfer efficiency of 21 $\%$ with $1~$MHz detuning and $\delta_{\text{FM}}$ = 1760 kHz, which is consistent with previous measurements on free-space red MOTs~\cite{Karori1999SrMOTPhotonRecoilTemp}. We characterize the BB MOT lifetime by observing the decay in atom number over time; see Fig. 2(b).  An exponential fit (dashed) indicates a lifetime of $(400 \pm 20)$ ms; the uncertainty represents 95$\%$ confidence intervals.

\begin{figure}[ht!]
\centering
\includegraphics[width=\linewidth,trim=7 2 1 1,clip]{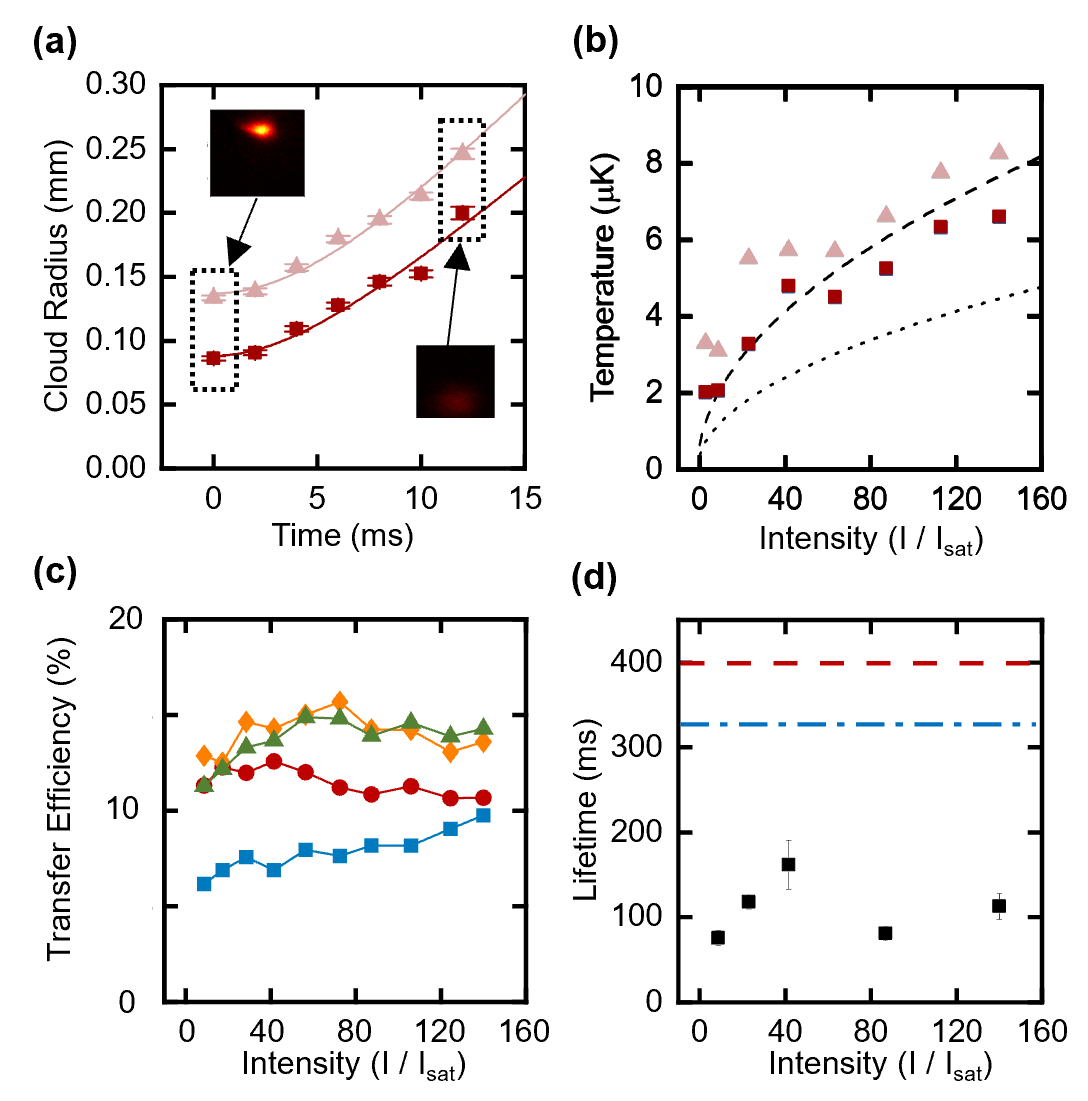}
\caption{SF MOT Characterization.  (a) Time-of-flight expansion of horizontal  (light red triangle)  and vertical (dark red square) cloud radius; Error bars represent 95$\%$ confidence intervals. (b)  SF MOT temperature with $I/I_{\text{sat}}$. Dashed lines indicated theoretical temperature $T = N_r~\times~\hbar~ \Gamma(I)/ 2 k_B  $ with $N_r = 2.1$ (dotted) and $N_r = 2.1\sqrt{3}$ (dashed). (c)  Transfer Efficiency with $I/I_{\text{sat}}$ for $\Delta_{\rm{final}} = -333~$ kHz (blue square), $\Delta_{\rm{final}} = -500~$ kHz (red circle), $\Delta_{\rm{final}} = -667~$ kHz (orange), and $\Delta_{\rm{final}} = -750~$ kHz (green) (d) SF MOT lifetime vs. intensity (black square); the blue MOT and BB MOT lifetimes are shown as a blue dash dot line and red dashed line respectively.
\label{fig:fig3}
}
\end{figure}

Figure \ref{fig:fig3} presents a characterization of the SF MOT at the conclusion of our experimental sequence. Here, we fix the red MOT parameters except for the intensity, which we vary to explore the low photon scattering regime that coincides with lowest temperature. We measure the temperature of the MOT with a time-of-flight technique, stroboscopically recording the atom-cloud size expansion after turning off the trap. At the end of the expansion time, a CMOS camera images the atom-cloud fluorescence from a $125~\mu$s pulse of 461 nm MOT light, collecting a column-integrated signal from the cloud. Figure~\ref{fig:fig3}~(a) presents such a measurement in which each data point corresponds to a variation in the free-expansion time and free-fall under gravity. 
Fitting the cloud radius $\sigma$ as a function of expansion time gives the MOT temperature through the relationship $\sigma(t) = \sqrt{\sigma(t = 0)^2 + k_B T / m \times t^2~}$, where $T$ is the temperature, and $m$ is the atomic mass. 
We characterize the temperature in the two observed directions, horizontal (light red triangle) and vertical (dark red square), since the non-conservative forces of a MOT do not completely thermalize the energy distribution of the trapped gas. 
Our system achieves average MOT temperature as low as $2~\mu$K, which is most likely cold enough to permit future experiments that load a magic-wavelength optical lattice with high efficiency. The insets show MOT images at the minimum and maximum expansion times of the dataset. Critically, our system produces atoms with microkelvin temperature, reasonable transfer efficiencies, and trapping lifetime that in principle enables loading into a magic-wavelength optical lattice for clock spectroscopy of the $^{1}$S$_0 \rightarrow ^{3}$P$_0$ transition. 

In the regime of large detuning relative to the power-broadened transition linewidth, the equilibrium of trapping forces of the MOT and Zeeman-induced detuning variation balance against gravity.
Thermodynamic modeling of narrow-line cooling predicts a SF MOT temperature  $T = N_r~\times~\hbar~ \Gamma(I)/ 2 k{_\text{B}}$  analogous to the Doppler limit, where  $\hbar$ is Planck's constant,  $\Gamma(I) = \Gamma~\sqrt{1 + I / I_{\text{sat}}}$ is the power broadened transition linewidth, $I$ is the single beam peak intensity, $I_{\text{sat}} = 3~ \mu$W/cm$^2$ is the saturation intensity of the transition, and $N_r$ is an overall scaling factor~\cite{Loftus2004NarrowCrystals}. In this regime of laser cooling, the final temperature of the MOT is dependent primarily on the laser intensity. Modeling of a traditional cubic MOT geometry, in which a single beam provides the force counteracting gravity, predicts a weakly intensity-dependent $N_r\approx2.1$~\cite{Loftus2004NarrowCrystals}.  

We systematically vary the SF MOT intensity to search for the lowest temperature; see Fig.~\ref{fig:fig3}~(b) with the black dotted line indicating the temperature predicted with $N_r~=~2.1$. The overall trend of the data indicates the intensity-dependent temperature of the SF MOT reaches approximately 2 $\mu$K  at $I / I_{\text{sat}}= 5$. However, in our system with three MOT beams oriented with vertical components, the temperature dependence with $\Gamma(I)$ is consistent with $\sqrt{3}$ larger laser power. Indeed, scaling the temperature theory according to $N_r=2.1\sqrt{3}$ is consistent with our observations; see the dashed line in Fig. 3~(b). 
These data highlight a tradespace of system design with integrated photonics, which largely constrains beam emission to a planar geometry, opening light-atom interactions to complexity from imperfect matching of polarization, magnetic fields~\cite{Leopold22_two_color_gmot}, the gravitational field, and geometrical imperfections that cause atoms to scatter light from additional MOT beams in the equilibrium position of a SF red MOT.
Indeed, our entire MS beam delivery system is assembled with no free-space optics and no adjustments. While we do not saturate the theoretical limit of achievable SF MOT temperature, our results are sufficient to load atoms into an optical lattice, which we will characterize in a future report.

In Fig.~\ref{fig:fig3}~(c), we explore the transfer efficiency from the blue MOT to the SF MOT as a function of the laser intensity and detuning at the conclusion of the experimental sequence. We measure the transfer efficiency for four final detunings $\Delta_{\rm{final}}$: $-333~$kHz (blue), $-500~$kHz (red), $-667~$kHz (orange), and $-750~$kHz (green). We observe a transfer efficiency between 11 $\%$ and 16 $\%$ without a significant intensity dependence, except for $\Delta_{\rm{final}} = -333~$kHz detuning. At this $\Delta_{\rm{final}}$, we see an approximately linear increase in transfer efficiency with intensity, peaking at 10 $\%$ at $I = 140~I_{\text{sat}}$. We measure the lifetime of the SF MOT to be between $80~$ms and $160~$ms; Fig.\ref{fig:fig3}~(d). Our blue MOT holds several million $^{88}$Sr atoms, hence the overall number and temperature we transfer to the SF MOT is promising to implement optical lattice loading and clock-transition spectroscopy.

In conclusion, our work demonstrates laser cooling of $^{88}$Sr atoms to microkelvin temperature, using an integrated-photonics system. Through the integration of metasurface optics on a fused silica substrate, we demonstrate both broad and narrow-line cooling free from bulk optics or alignment of individual MOT beams. Our system employs fiber-packaged, integrated nonlinear waveguides to spectrally broaden a frequency comb, enabling frequency stabilization and linewidth narrowing of the cooling lasers. Our findings eliminate barriers to employing ultracold atomic physics systems.  
\\

This work was supported by DARPA A-PhI FA9453-19-C-0029, AFOSR FA9550-20-1-0004 Project Number 19RT1019, NSF Quantum Leap Challenge Institute Award OMA - 2016244, and NIST. This work is a contribution of the U.S. government and is not subject to copyright. Trade names provide information only and not an endorsement.

\bibliography{AMO_references}

\end{document}